\documentclass[sigconf]{acmart}

\usepackage{dirtytalk}
\usepackage[colorinlistoftodos]{todonotes}
\usepackage{textcomp}
\usepackage{comment}
\usepackage{quoting}
\quotingsetup{vskip=10pt}

\AtBeginDocument{%
  }

\copyrightyear{2022} 
\acmYear{2022} 
\setcopyright{acmcopyright}\acmConference[ESEM '22]{ACM / IEEE International Symposium on Empirical Software Engineering and Measurement (ESEM)}{September 19--23, 2022}{Helsinki, Finland}
\acmBooktitle{ACM / IEEE International Symposium on Empirical Software Engineering and Measurement (ESEM) (ESEM '22), September 19--23, 2022, Helsinki, Finland}
\acmPrice{15.00}
\acmDOI{10.1145/3544902.3546234}
\acmISBN{978-1-4503-9427-7/22/09}

\begin{document}

\title{Understanding the Implementation of Technical Measures in the Process of Data Privacy Compliance: A Qualitative Study}

\author{Oleksandra Klymenko}
\affiliation{%
 \institution{Technical University of Munich}
 \streetaddress{Boltzmannstr. 3}
 \city{Garching}
 \state{Bavaria}
 \country{Germany}
 \postcode{85748}
}
\email{alexandra.klymenko@tum.de}

\author{Oleksandr Kosenkov}
\affiliation{%
 \institution{fortiss GmbH}
 \streetaddress{Guerickestr. 25}
 \city{Munich}
 \state{Bavaria}
 \country{Germany}
 \postcode{80805}
}
\email{kosenkov@fortiss.org}

\author{Stephen Meisenbacher}
\affiliation{%
 \institution{Technical University of Munich}
 \streetaddress{Boltzmannstr. 3}
 \city{Garching}
 \state{Bavaria}
 \country{Germany}
 \postcode{85748}
}
\email{stephen.meisenbacher@tum.de}

\author{Parisa Elahidoost}
\affiliation{%
 \institution{fortiss GmbH}
 \streetaddress{Guerickestr. 25}
 \city{Munich}
 \state{Bavaria}
 \country{Germany}
 \postcode{80805}
}
\email{elahidoost@fortiss.org}

\author{Daniel Mendez}
\affiliation{%
 \institution{Blekinge Institute of Technology}
 \streetaddress{Valhallavägen 1}
 \city{Karlskrona}
 \country{Sweden}
 \postcode{371 41}
}
\email{daniel.mendez@bth.se}

\author{Florian Matthes}
\affiliation{%
 \institution{Technical University of Munich}
 \streetaddress{Boltzmannstr. 3}
 \city{Garching}
 \state{Bavaria}
 \country{Germany}
 \postcode{85748}
}
\email{matthes@tum.de}

\renewcommand{\shortauthors}{Klymenko et al.}

\begin{abstract}
\textbf{Background}: Modern privacy regulations, such as the General Data Protection Regulation (GDPR), address privacy in software systems in a technologically agnostic way by mentioning general "technical measures" for data privacy compliance rather than dictating how these should be implemented. An understanding of the concept of technical measures and how exactly these can be handled in practice, however, is not trivial due to its interdisciplinary nature and the necessary technical-legal interactions.


\textbf{Aims}: We aim to investigate how the concept of technical measures for data privacy compliance is understood in practice as well as the technical-legal interaction intrinsic to the process of implementing those technical measures.

\textbf{Methods}: We follow a research design that is 1) exploratory in nature, 2) qualitative, and 3) interview-based, with 16 selected privacy professionals in the technical and legal domains.

\textbf{Results}: Our results suggest that there is no clear mutual understanding and commonly accepted approach to handling technical measures. Both technical and legal roles are involved in the implementation of such measures. While they still often operate in separate spheres, a predominant opinion amongst the interviewees is to promote more interdisciplinary collaboration.

\textbf{Conclusions}:
Our empirical findings confirm the need for better interaction between legal and engineering teams when implementing technical measures for data privacy. We posit that interdisciplinary collaboration is paramount to a more complete understanding of technical measures, which currently lacks a mutually accepted notion. Yet, as strongly suggested by our results, there is still a lack of systematic approaches to such interaction. Therefore, the results strengthen our confidence in the need for further investigations into the technical-legal dynamic of data privacy compliance.

\end{abstract}



\maketitle

\section{Introduction}
In recent years, regulators around the world have been active in enacting new privacy regulations, mainly motivated by the rapid advancements in software systems and their data processing capacities. Still, many regulations such as GDPR, the California Consumer Privacy Act (CCPA), and the older Healthcare Insurance Portability and Accountability Act (HIPAA) are "technology neutral", in the sense that they abstract away from the technology to be regulated \cite{schmidt_should_2006}. Hence, these regulations do not prescribe explicit requirements for the implementation of specific technologies \cite{hhs_standards_2010} that would render the final software-intensive products compliant. The rationale for technology neutrality is to allow for flexibility of regulatory frameworks and to accommodate future technological developments. This comes at the price of the absence of concrete specifications for the implementation of regulatory norms via "technical measures" (i.e. in GDPR).

The handling of technical measures in practice, i.e. their interpretation, translation to requirements, and their implementation, can raise various challenges. For example, according to Smith et al. \citep{smith_analysis_2021}, three years after the GDPR came into effect, one of the main GDPR-related challenges of day-to-day operations in organizations is the implementation of requirements into multiple systems. In GDPR, the term "technical measures" is written into the letter of the law as an umbrella term that covers multiple measures and principles.
Such technical measures are dispersed in the text of GDPR and can be identified directly (for example, the principles of data protection by design or the need for pseudonymization). Huth et al. \cite{Huth2019AppropriateTA} identified 8 such technical measures and their properties in GDPR and established their definitions based on GDPR, ISO 27000:2018, and research publications on privacy terminology. Still, the identification of such measures as shown in \cite{Huth2019AppropriateTA} requires at least profound knowledge of GDPR and auxiliary relevant sources.

On the other hand, technical measures need to be identified and implemented even if GDPR does not expound upon them directly. Regulators emphasized this by using the term \say{appropriate} in relation to technical measures. Still, deciphering "appropriateness" alone can require an academic level legal inquiry \cite{selzer_practitioners_2021}. In this light, it is reasonable to assume that understanding technical measures for data privacy compliance requires at least domain knowledge and legal expertise in software engineering teams. Available empirical research in the domain of regulatory compliance of software systems has shown that software engineers cannot independently cope with complexities, and legal experts' support can be important \cite{maxwell2013empirical}. Other research in the data privacy domain also suggested that the involvement of legal experts is often required whereas the interaction between both legal experts and software engineering teams is often cumbersome \cite{usman2020compliance} and challenging \cite{zimmermann2020automation}.

The importance of such technical-legal interaction cannot be underestimated, as it heavily influences the design of a software system and eventually dictates its compliance (assurance). While the research community has already pointed to technical-legal interactions as an important aspect of technical measures implementation, little is yet known about the way in which technical measures are really understood and implemented in practice. The lack of understanding of this interaction may result in inefficient practices.

The investigation of the technical-legal interaction is challenging as it requires one to consider both engineering and legal perspectives. This investigation represents an important research gap, as pointed out by Altman et al. \cite{Altman2020}, namely that \say{legal and technical approaches to data protection have developed in parallel, and their conceptual underpinnings are growing increasingly divergent}. 

In this paper, we contribute an interview study to explore legal and engineering practitioners’ perspective on technical measures for data privacy as well as their implementation in practice with a specific focus on the interaction between legal and engineering roles. We deliberately chose to focus on "technical measures" as this term enables systematic investigation and integration of both technical and legal perspectives. 

As such, we first aim to investigate the current understanding of the processes that take place in order to implement technical measures for data privacy compliance in practice. In supplement, we explore the structures existing to facilitate such processes, particularly the roles and interactions therein. We base this approach upon the hypothesis that in order to begin to understand \say{technical measures}, one must first look behind the curtains to the ecosystem of roles, interactions, and decisions defined and existing within the process of data privacy compliance.

In the following, we first lay the terminological foundation (Section~\ref{sec:fundamentals}) and discuss work related to our research (Section~\ref{sec:relatedwork}) before introducing the followed research methodology in Section~\ref{sec:researchmethodology}. In Section~\ref{sec:results}, we present our results and critically reflect upon them in Section~\ref{discuss}. In Section~\ref{sec:threats}, we discuss selected threats to validity before concluding our paper in Section~\ref{sec:conclusion}.

\section{Fundamentals and Background}
\label{sec:fundamentals}


\subsection{Regulation and Data Privacy Compliance}
When one looks at the legal side of privacy, the response has largely come in the form of regulations. In some places, it is even named a fundamental right by law. The understanding that privacy is a fundamental right is not novel, yet earlier notions of this right evolve around the \say{limitation of governmental power} \cite{rubenfeld1989right}. With modern regulations, the discussion has shifted to that of \textit{data privacy}.

The General Data Protection Regulation\footnote{\url{https://eur-lex.europa.eu/legal-content/EN/TXT/PDF/?uri=CELEX:32016R0679}}, often shortened to GDPR, is touted as \say{the toughest privacy and security law in the world}, having implications far beyond the borders of the European Union as it applies to any entity collecting or processing data that originates from within the EU. Goddard \cite{Goddard2017} reinforces this notion of the \say{wide jurisdictional scope} of GDPR. In this paper, we directly focus on one of its most interesting aspects, and arguably one that makes GDPR so relevant - \textit{Data Privacy Compliance}.

Of course, compliance of software-intensive products to legal regulation is nothing new. The novelty introduced by GDPR comes firstly with the widespread applicability and enforceability resulting from its nature as an EU-wide regulation. This concerns the protection of data privacy in an age of increasing data utilization, which certainly amounted to quite the change. Such novel scope is captured in works such as de Hert et al. \cite{deHert2016}.

As such, GDPR introduced a new urgency to the demonstration of privacy compliance, with the possibility of rather significant fines in the event of non-compliance. One study by Wolff et al. \cite{Wolff2021} of GDPR fines up until 2019 found that \say{the largest volume of fines... were levied due to violations in which organizations implemented “insufficient technical and organizational measures to ensure information security,” totaling some 59 fines that amounted to €332,864,417}. This highlights the new pressure placed on data processors to implement the appropriate \say{technical measures}. Recent works such as Chhetri et al. \cite{Chhetri2022} highlight the persistent challenges of compliance verification at scale and interoperability introduced by the \say{paradigm shift in data protection} created by GDPR.

It is important to note that GDPR is by no means the only relevant regulation. In an interesting and possibly foreseeable way, many of the newest regulations around the globe certainly take inspiration \say{in the wake of GDPR} \cite{greenleaf2019global}, further pointing to the monumental impact of this single regulation.

\subsection{Technical Measures}
The idea of \say{technical measures} is clearly central to the work presented here, yet it is nevertheless a bit unclear what exactly might be meant by this. The term is quite unambiguous in the sense that technical measures call for ways in which technology can be used for the purposes of data protection. Beyond that, though, technical measures can presumably take many forms, yet the general understanding of such measures remains vague and is not yet well explored by the research community. 

There has been work, mostly on the level of national supervisory authorities, to attempt to clear up the technical side of privacy compliance via the release of overarching \say{guidelines}. Those guidelines provide a valuable first step towards privacy compliance. More importantly, and something that may lie at the core of some challenges to be explored in this paper, the text of the German Standard Data Protection Model (SDM)\footnote{\url{https://www.datenschutzzentrum.de/uploads/sdm/SDM-Methodology_V2.0b.pdf}} acknowledges that regulation cannot be:

\begin{quote}
"...readily operationalised in a technical manner. Lawyers and computer scientists must therefore find \textbf{a common language} to ensure that the legal requirements are actually implemented technically."
\end{quote}

Despite this common understanding, the technical measures for data privacy compliance are not covered thoroughly in the available guidelines. Exploring the crux of this statement becomes the focal point of this work, lying at the intersection of technical and legal perspectives on privacy compliance.

\section{Related Work}
\label{sec:relatedwork}
Software engineering research related to the implementation of technical measures for data privacy compliance can be mainly found in the fields of privacy engineering and regulatory requirements engineering, both of which are inherently interdisciplinary \cite{gurses2016privacy, kosenkov2021vision}.

There are discrepancies in the research focusing on privacy in a broad sense and compliance to GDPR or other privacy regulations. Usually, privacy engineering research has a broader scope focusing on the development of "privacy friendly systems" that among other goals includes the fulfilment of privacy-related user expectations \cite{Spiekermann2009}, rather than simply achieving regulatory compliance.

Zimmermann~\cite{zimmermann2020automation} suggested a privacy engineering reference process as a basis for automation. The reference model includes legal tasks (e.g., identification of applicable regulations, etc.), but there is no separate explicit category of legal tasks. The author mentions that such legal tasks require interpretation and case-specific balancing. That makes them less suitable candidates for full automation. The author also does not further elaborate on the differentiation between engineering and legal tasks, the way interaction should be organized, or the roles that should be involved.


There also exist various empirical studies related to regulatory compliance of software systems overall and to GDPR in particular. For instance, Massey et al. \cite{massey2015strategy} investigated ambiguities in regulatory norms and suggested a strategy to address them. They found that while software engineers can successfully identify such ambiguities, they still require further support or legal expertise to resolve them. The authors also suggested that the division of labor is required to address legal ambiguities effectively. Organization of such division of labor was not in the scope of their work. Maxwel et al. \cite{maxwell2013empirical} also investigated the ability of software engineers to address legal cross-references, and they concluded that engineers are ill equipped to do it without any legal expert support.

Sirur et al. \cite{sirur2018we} conducted an interview study in order to reveal challenges in compliance with the GDPR. Their results showed that more technically focused respondents expressed the opinion that the average engineer would struggle to apply the regulations directly without support from a legal expert. These findings are well aligned with the empirical research by Alhazmi et al. \cite{alhazmi2021m} on the implementation of GDPR by software engineers. In their study, they found that the majority of software engineers seem to believe that data privacy compliance was not their responsibility and some of them expressed an opinion that there should be a separate team responsible for data privacy.


Finally, while works such as by Altman et al. \cite{Altman2020} acknowledge the divergence between technical and legal perspectives on privacy compliance, they do not place a particular focus on technical measures as the junction point. Furthermore, the process, roles, and interactions surrounding technical measures are likewise not addressed. Similarly, in the work of Piras et al. \cite{piras2019defend}, the authors support the notion that the GDPR lacks \say{providing details concerning technical and organizational privacy and security measures needed, and therefore it is difficult for organizations to be fully GDPR compliant}.

As shown, multiple publications highlight the need for technical-legal interaction. Still, there is no one systematic understanding of it based on the practice of the implementation of technical measures for data privacy compliance.

\section{Research Methodology}
\label{sec:researchmethodology}
In this section, we introduce the research methodology followed in our interview study.
\subsection{Goal and Research Questions} \label{goal}
In our study, we aim to explore the implementation of technical measures for data privacy compliance in practice, from the perspective of both legal and engineering professionals. A specific focus is on the interaction between legal and engineering roles.


From our goal, we infer three major research questions (RQs):
\begin{enumerate}
    \item[RQ 1] How are technical measures for data privacy compliance understood in practice?  
    \item[RQ 2] Which roles and responsibilities are involved in the implementation of technical measures? 
    \item[RQ 3] What interactions are shared in practice between the technical and legal proponents of data privacy compliance?
\end{enumerate}

The first question aims to understand practitioners' perceptions of \say{technical measures} in their individual organizational context. The second research question is then geared towards the processes put in place for the implementation of technical measures, and the roles that are defined therein. Finally, under the umbrella of the third research question, we aim to explore the interactions that take place in these processes. As a result, we hope not only to gain a static picture of what technical measures are understood to be, but also to understand the dynamics taking place when handling technical measures in software engineering processes.

\subsection{Research Process}
We follow a qualitative research approach by conducting semi-structured interviews, thus drawing from the experiences and expertise of participants working with privacy in either the legal or technical sectors, or ideally ones that traverse both fields. For this study, we draw from the Grounded Theory methodology as described by Hoda et al.~\cite{GT11}. In particular, we followed the approach of first presenting a pre-defined set of questions to our interview participants, with the goal of understanding their roles, responsibilities, interactions, and future goals regarding privacy compliance within their organizations and/or fields. We then recorded and transcribed the interviews, coding and constantly comparing the results of each interview, until we reached a saturation that allowed us to close the interview study. We conducted a total of 16 interviews.

Although the interviewees allowed for the findings to be published, they did not agree to disclose the full transcriptions. Therefore, in the interest of privacy, these are not included. 

\subsubsection{Identifying Participants}
For the process of identifying interviewees, the source of contacts was limited to four avenues:
\begin{itemize}
    \itemsep 0em
    \item\textit{Personal contacts}: people who could be immediately contacted as a result of a shared connection
    \item \textit{Pre-saved contacts}: i.e. from the work of previous research, in which contact lists were created
    \item \textit{Top search results}: i.e. via LinkedIn, where terms such as 'privacy professional', 'privacy engineer', 'data protection officer', or 'data privacy lawyer' were entered
    \item \textit{Referrals}: once initial interviews were conducted, people referred to within these could then be contacted, now with a shared point of reference
\end{itemize}

Once these contacts were identified, several steps were followed to initiate contact, present the opportunity for an interview, and schedule the interview. The basic process is as follows:
\begin{enumerate}
    \itemsep 0em
    \item Informally ask for an interview (e.g. LinkedIn direct message)
    \item Follow-up/send a formal email invitation, outlining the purpose of the research and structure of the interview
    \item Schedule the interview
    \item Before the interview, provide the set of questions so that the interviewee can prepare his or her responses
\end{enumerate}
Particularly with the last point, this step was seen as necessary due to the nature of the semi-structured interviews. Since the interviewee was provided with the questions beforehand, the general flow of the interview would be known in advance, allowing for impromptu follow-up questions and deeper discussions.

This approach for participant selection was viewed to be in line with the goal to obtain different perspectives from privacy professionals, both legal and technical. By searching for roles preconceived to be involved with the compliance process, the interview pool could be molded in an opportunistic, yet strategic manner.

\subsubsection{Interviewee Demographics}

Table \ref{tab:coded_interviewee} presents a codified table of our sample including the interview participants' relevant information, such as unique interviewee codes that will be referenced throughout in order to cite specific statements or express a particular opinion held by the respective interviewee. Codes suffixed with a 'T' denote a technical contact, 'L' a legal, and 'LT' a technical/legal contact.
\say{Exp.} denotes years of experience. In cases where the interviewee has longer relevant experience in the technical or legal fields, and later on began working as a privacy professional, the number of years worked as a \textit{privacy professional} is indicated in the parentheses. \say{Dur.} denotes the duration of the interview.

The interviews consisted of 11 male and 5 female interviewees, from organizations spanning three continents. Another pertinent demographic is the 154 cumulative years of experience shared amongst the interviewees: 76 of these come pre-GDPR, and the remaining 78 post-GDPR. This is interesting to note as the perception of technical measures might very well be influenced by viewpoints taken prior to the adoption of GDPR while the influence of GDPR will most certainly also be heavily reflected.

\begin{table*}[hbtp]
  \caption{Interview Study Participants}
  \vspace{-10pt}
  \centering
    \begin{tabular}{p{0.085\linewidth} | p{0.3\linewidth} | p{0.4\linewidth} | p{0.05\linewidth} | p{0.05\linewidth}}
    \toprule
    {\textbf{Participant ID}} & {\textbf{Position}} & {\textbf{Organization}} & {\textbf{Exp.}} & {\textbf{Dur.}} \\
    \midrule
    I1-T  & Privacy Engineer & Large US media conglomerate & 10+ (1) & 54 \\
    \midrule
    I2-T  & Privacy/Security Architect & Large German multinational software corporation & 6 & 52 \\
    \midrule
    I3-L  & Privacy and cybersecurity lawyer & US law firm & 20+   & 32 \\
    \midrule
    I4-T  & Privacy Engineer & Large US multinational tech company & 5+ (4) & 70 \\
    \midrule
    I5-LT & DPO, Managing Director & Small German-based data protection software company & 4 & 50 \\
    \midrule
    I6-T  & Software Architect & Large German multinational tech conglomerate & 3 & 50 \\
    \midrule
    I7-L  & Lawyer/external DPO & Small German data privacy company & 20+  & 55 \\
    \midrule
    I8-L  & Group Data Protection Counsel & International financial technology corporation & 6 & 60 \\
    \midrule
    I9-T  & Privacy Engineer & Large US Tech Corporation & 8 & 65 \\
    \midrule
    I10-LT & Legal Counsel & Global Web Consortium & 25 & 60 \\
    \midrule
    I11-L & Legal Counsel & German-based digital privacy consulting firm & 3 & 50 \\
    \midrule
    I12-L & DPO  & German-based consulting firm & 20 (3) & 55 \\
    \midrule
    I13-T & Security and Privacy Architect & Large German multinational tech conglomerate & 3 & 60 \\
    \midrule
    I14-L & Compliance Officer & British-based news corporation & 3 & 50 \\
    \midrule
    I15-T & Privacy Engineer & Chinese multinational tech corporation & 15 & 60 \\
    \midrule
    I16-L & Legal Associate & Indian-based law firm & 3 & 55 \\
    \bottomrule
    \end{tabular}%
  \label{tab:coded_interviewee}
\end{table*}%

\subsubsection{Instrumentation}

We developed an interview guideline based on the research questions provided in Section \ref{goal}. The guideline consists of different sections: First, \textit{general questions} about the participants' backgrounds are considered. The next section includes questions that inquire about the process of \textit{regulation identification and interpretation}, based on the interviewees' experiences and understanding. Furthermore, our participants were asked to discuss the strategies and tools that they use for the implementation of the identified privacy regulation. The final part investigates \textit{recommendations and future directions}.

\subsubsection{Coding Process}
The results of the interviews were coded in accordance with the guidelines of a Thematic Content Analysis \cite{braun2012thematic}. In our data analysis, we followed the ensuing steps:
\\ \\
\begin{minipage}{8cm}
\begin{enumerate}
    \itemsep 0em
    \item Transcript reading – reviewing the conducted interview
    \item Transcript annotation – highlighting important words, \\ phrases, points, etc. in the raw transcripts
    \item Data conceptualization – creating common themes and codes based upon the annotated transcript data
    \item Data segmentation – marking the transcripts according to category (each highlighted segment receives a code)
    \item Verification – validating if the themes accurately depict the transcript data
    \item Analysis and Results – writing a summary of the interview with the help of the annotated transcripts and its themes
\end{enumerate}
\end{minipage}

\section{Study Results}
\label{sec:results}
In the following, we present the results of our study, structured according to the research questions. 

\subsection{Practical Understanding of "Technical Measures" (RQ~1)}
\label{rq1}
In the interviews, the approach towards answering our first research question was carried out by posing the following question to the interviewees: \textit{What is your understanding of the “technical measures” required to comply with privacy regulations?} The findings from the responses to this question are presented, and above all, they call for an interesting analysis, not only in the content itself, but also by placing the responses in juxtaposition. Concretely, the foremost striking characteristic of the responses is the clear lack of uniformity. These various directions are introduced in the following.

The idea of technical measures as a risk assessment was reinforced quite often by the interviewees. I3-L provides a clear definition of technical measures as the \say{methods and processes we are going to put in place, from an organizational corporate standpoint, to reduce risk}. Interestingly, such a definition strays away from the technical aspect of the term in question, instead focusing on measures taken to address risk from an organizational viewpoint. Another interviewee, I7-L, poses the flip side of the coin, claiming that one is \say{taking risks} if the correct technical measures are not in place. Combining these two ideas, I9-T talks of using technical measures as a way to \say{mitigate privacy risks}. In this light, one may understand technical measures to be actions that serve to mitigate privacy risks, according to the risk assessment (and tolerance) of a particular organization.

Of course, even this risk assessment may differ greatly amongst organizations. An interesting factor relating to this notion is brought to light by I3-L, who emphasizes the cultural factor of risk assessments when it comes to technical measures. Essentially, there are two fields of thought on the matter. The first abstracts from the rigidity of a term like \textit{technical measures}, and essentially boils privacy compliance down to risk reduction. The other perspective, following the \say{letter of the law}, indeed focuses on technical measures, for \say{if the law says that we should implement technical measures, we will implement technical measures} (I3-L). In the end, the distinction lies within the decision over whether technical measures must be exactly that, or rather if technical (and organizational) measures represent a matter of interpretation, or \say{degrees of freedom} (I2-T).

To the latter point of view, responses to the question also focused on the purely technical aspect of the technical measures in question. I5-LT mentions that \say{hardware or software installations} must be put into place in order to prevent privacy breaches. On the consulting side, there may often be checklists utilized in order to understand the technical systems in play, so as to begin to tackle the implementation of technical measures. For some interviewees, technical measures go hand in hand with encryption, which certainly would fall under the umbrella of the term. Interestingly, the frequent mention of encryption may not come as a surprise, for it counts as one of the rare cases of technologies explicitly mentioned in the text of GDPR. Likewise, I16-L makes mention of anonymization (also in GDPR), and I12-L points to other measures such as a data intrusion detection system. As one may extract from such responses, the idea of technical measures is quite varied, even when confined to the strictly technical sense.

This non-uniformity comes as a particular issue for some interviewees, who see this flexibility as a challenge. I2-L is sometimes perplexed by \say{multiple options} for technical measures, or rather, the \say{it depends} scenarios. This is echoed by I6-T, who views the implementation of technical measures as a \say{broad question, and it has multiple layers of answers to it}. From a legal perspective, the lack of clarity on technical measures \say{creates a lot of issues and a lot of problems and a lot of headaches} (I8-L). Beyond this, I8-L believes that technical measures include other actions \say{that might not be listed explicitly as sort of technical measures but are good on the regulation side}, further muddying the waters in the understanding of technical measures.

The legal perspective provides a hint of clarity on the matter, with regards to the discussion of \textit{data flow}. A simple set of questions becomes: \say{What do you do with the data? What kind of data do you have?} (I8-L). In this light, the securing of data becomes a focal point of technical measures. Another definition of technical measures comes from I16-L as \say{multiple measures to be taken by the organization as far as like what type of practices they do with the data}, thus enabling the organization to \say{define the data flow}. With this, though, I8-L reminds us that \say{there will always be different data elements, different kinds of sensitivity of the data, different infrastructures in place}, which forces the notion of technical measures to retain in part its flexible nature.

Of particular interest to the technical aspect is the question of how Privacy-Enhancing Technologies (PETs) fit into the discussion of technical measures for privacy compliance. I2-T references specific examples such as Homomorphic Encryption and Differential Privacy, yet the challenge there becomes whether one can \say{claim to have any benefit}, in regards to the demonstration of compliance (and avoidance of fines), from using \say{the most sound techniques}, i.e. PETs. This highlights the issue that although PETs may be technically sound in regards to privacy preservation, their adoption for compliance purposes is unclear. In the same discussion, I2-T makes this issue concrete by saying \say{the mapping is not yet very clear}. Another viewpoint on this matter comes from I4-T, who says that the decision to implement PETs comes from \say{high up}, implying that the decision to interpret PETs as appropriate technical measures is one left to upper management. A final insight into the understanding of PETs of technical measures comes from I9-T, who believes that many PETs are currently \say{too academic} to be applicable as technical measures. This rehashes the importance of education and awareness in enabling the adoption of PETs in real-world scenarios, that is, outside of the academic sphere.

In a similar way, the significance of automation in the process of data privacy compliance is something that was inquired about. However, the collective response was not entirely thorough, often amounting to agreement with the fact that automation could be helpful in the process, but is currently not being utilized. 

Another important aspect to the discussion of technical measures comes with a retort from the technical viewpoint, expressed with the statement: \say{there's a very big difference between a legal analysis or assessment and a technical privacy implementation} (I9-T). This is echoed and even complicated a bit by I9-T: \say{It is even more important than just the technical}. Thus, in the conversation on the understanding of technical measures and their implementation for privacy compliance, the scope is expanded to external factors. The statement by I9-T suggests that the discrepancy between legal mandate and technical implementation may be complicating the current understanding of \say{technical measures} for legal compliance. Even within the technical sphere, I4-T reminds us that although a technical measure may in some cases be relatively simple to \textit{implement}, there are complicating factors in the maintenance of such measures, including the tracking of data flow.

Parallel to the understanding of technical measures in practice, the larger scope of the privacy compliance process must also be taken into consideration. In particular, the concept of \textit{enforceability} becomes crucial, per I10-LT. Requiring technical measures to take \say{due regard to the state of the art} (GDPR) is not only inherently vague, but it may also lead organizations \say{to unproportionally invest to achieve compliance} (I10-LT). This problem is only exacerbated if the enforcement of data privacy compliance is not clear. Thus, if the \say{rule of law is not existing, we are in this kind of foggy environment} (I10-LT). This without a doubt plays a significant role in the perception of technical measures. 

The inquiry into a practical understanding of technical measures for privacy compliance certainly elicited rich and varied responses. As such, the further discussion of the findings introduced here is merited (Section \ref{discuss}). Before this, we explore the context surrounding this discussion, namely the role and interactions involved in the implementation of technical measures for privacy compliance.

\vspace{5pt}
\fbox{
\begin{minipage}{24em}
RQ1: The inquiry into understanding technical measures elicited varied responses, pointing to their complex nature.
\end{minipage}}

\subsection{Structure of the Technical Measures Implementation Process (RQ 2)}
\label{rq2}
In order to understand the process behind the implementation of technical measures to enable data privacy compliance, it is first necessary to identify the involved parties and their individual responsibilities. Through the interviews, we extracted these based on the interviewees' descriptions of their roles and responsibilities as well as their interactions with other roles involved in the process of data privacy compliance (the latter is covered in Section \ref{interactions}).

\vspace{1pt}

\vspace{-5pt}
\subsubsection{Legal side} The first party involved on the legal side of the process is the \textit{Legal Team}, or \textit{Legal Counsel} -- a group of practicing lawyers and/or legal associates specializing in sub-fields such as privacy, data protection and cybersecurity. Larger organizations might have \textit{in-house counsel} and internal legal team(s), while smaller organizations or ones where the constant availability of legal support is not necessarily needed often use \textit{outside / external counsel}. The responsibility of the legal counsel is to provide interpretation of the relevant regulations to other (non-legal) parties to develop and maintain compliant systems: \say{Legal interprets everything… their job is to make sure we're staying up to date, trying to be compliant.} (I1-T). The job of outside counsel is virtually identical: \say{We're outside counsel, which means that we advise clients on the legal requirements.} (I3-L). Again, this emphasizes the advisory nature of legal support, that is the guidance provided by lawyers as the interpreters of laws and regulations.

Also under the legal category but strictly distinct from legal counsel is the role of an \textit{External Consultant}. This person is most often involved with small and mid-sized organizations and is \say{specialized in providing support on privacy topics} (I7-L), especially including compliance. Consultants \say{support and consult with companies that want to process data of individuals… to make sure that the client is really handling the data in a legal matter, and in a secure space} (I11-L).
Crucial to note is that for external consultants there is no expectation that the consultant is a practicing lawyer. Furthermore, this rarely seems to be the case; of the interviewees working in this role, none were lawyers. From this fact comes the true distinction between somewhat overlapping roles of legal counsel and external consultant -- under the eyes of the law, legal advice provided by a practicing lawyer cannot be replaced by that of a consultant.

The role of \textit{Compliance Officer} is quite unique in the sense that it is not a strictly legal role, rather, \say{It's a bit more on the legal side of it, without stepping too far into legal because we do have a legal team... we're simply there for the operational risk, and really deciding what that risk is meant to look like for us.} (I14-L). This explanation introduces the extremely vital concept of \textit{"go-betweens"} in the data privacy compliance structure, something that becomes very important and that will be expanded upon below. As is made clear by the title, the compliance officer is essentially involved with compliance-related matters within an organization. In the words of one such officer, \say{A lot of what we do is obviously ensuring that data privacy alone, amongst other compliance areas are [followed] for the rest of business… privacy obviously being the larger one of the pillars.} (I14-L). Compliance officers serve an integral role in proper compliance, although it is not clear how widespread this role is, i.e. in how many organizations such a role exists.

\subsubsection{Technical side} Starting off with more of an umbrella term, the concept of the \textit{Product Team} encompasses the team involved in the design and implementation of \textit{the product}. In the technical sense, this usually refers to the system(s) being developed for the market. Specifically on the product team may be roles such as (software) developers and engineers, as well as the appropriate leadership roles. The product team, lead by the project owner, is connected with the necessary officers to perform a risk analysis of the proposed project, which then serves as the basis for determining the necessary measures for compliance. The role of the project owner is also emphasized as \say{the first point of contact [regarding privacy]... it's the project owner's responsibility to deal with communication and bringing people together} (I6-T). Through this, one can see a hierarchy being built with regards to privacy matters within a technical vertical.

One specific role hailing from the technical vertical is that of the \textit{architect}. Using the more general term of \textit{Software Architect}, this person is more involved on the design side, rather than development or implementation. For this reason, the architect's role becomes very important in the implementation of compliant systems, as it is in this stage of (pre-)development that sound privacy-respecting practices are planned and incorporated. The role of Software Architect is sometimes specialized to \textit{Privacy Architect}, or also \textit{Privacy and Security Architect}, or even \textit{Enterprise Architect} (more generalized) in larger organizations. With this, the emphasis on the design of privacy-preserving, secure systems is made concrete.

The role of \textit{Management}, although not technical per se, is vital in the privacy compliance structure. The concept of management first and foremost includes the \say{C-Suite} executives responsible for the leadership and direction of organizations as a whole. It is often the case that \say{decisions making [regarding compliance] come from a couple levels above} (I1-T), i.e. management. A specific role that came up in multiple interviews is the (Chief) Information Security Officer. Interestingly enough, the title is misleading in the sense that the responsibility for \textit{privacy} matters also falls under this role as well. Put in rather general terms, \say{The ISec officer is responsible for ensuring there are no loopholes} (I6-T) in the development of compliant systems.

\subsubsection{The Go-Betweens} The next category, called the \textit{Go-Betweens} (or \textit{in-betweens}), of the privacy compliance structure has been defined by the authors to indicate the inherent interdisciplinary responsibilities possessed by these roles, often transcending one single field or sector.

The role of the \textit{Data Protection Officer (DPO)} has become central to the privacy compliance process. The general work of a DPO includes aiding in the compliance process for relevant data protection laws, accomplished via \say{monitoring specific processes, such as data protection impact assessments or the awareness-raising and training of employees for data protection, as well as collaborating with the supervisory authorities}\footnote{https://gdpr-info.eu/issues/data-protection-officer/}. The appointment of a DPO is not a blanket requirement, but rather depends on the data processing activities of the organization. Companies are also allowed relative freedom with regards to selection of a DPO, having the flexibility in the choice of an internal or external DPO.
The DPOs are "de facto go-betweens" (I3-L), and in the words of an acting DPO, "The person taking care of data protection in an organization needs to be hybrid." (I5-LT). At the core of the responsibilities of this person lies the task of liaising between what is said in the law, i.e. the \textit{letter of the law}, and how this is implemented in practice. Ultimately, it is crucial to note that, \say{At the end of the day, the data protection officer is more a legal person than a technical person.} (I2-T).

Another more hybrid, in-between role is that of a \textit{Privacy Engineer}, who is also placed at the intersection of law and technology, but is more on the technical side of the process. As defined by one of the interviewees, \say{Privacy engineering is really about ensuring there is the trust at the level of technology to protect privacy and to mitigate privacy risks.} (I9-T). A major responsibility of a privacy engineer comes as a policy maker for an organization: \say{We're the ones who define, dictate, and do privacy assessments.} (I9-T). It is interesting to note that \say{you more or less get very different backgrounds in privacy engineering} (I4-T). Privacy engineers do not necessarily need to be experts in the finer details of the technologies and implementations themselves; it is their expertise in data protection, privacy design principles, and the implications of privacy to society at large that provide an unique supplement to the design and implementation of technology, making the role so valuable, and at the same time challenging. As profoundly described by I4-T: \say{Privacy engineering is fundamentally dealing with science that has not yet been codified.}

\subsubsection{External stakeholders}
In the process of privacy compliance, much of the work when it comes to tooling and automation is often outsourced to \textit{Third Party Vendors}. These vendors often comprise of external companies providing a service or technology as a \say{technical measure} for compliance. One interviewee confirms that \say{most of the external interactions we have is with third party software providers for security functions and when it's about tooling.} (I2-T). This solution is useful for organizations looking for an \say{out-of-the-box} system, yet the introduction of third parties into a process centered on privacy can also raise concerns and/or challenges.

\textit{Supervisory Authorities} can serve as a useful resource for organizations in guiding privacy compliance programs. It is interesting to note that while several interviewees mentioned them only in passing, a clear need for more interaction was often expressed. 

The final external actor, arguably the most important, is the \textit{Customer}. While this stakeholder was not often mentioned explicitly, the implicit inclusion of this person was without a doubt always tacitly understood. One must not forget that in the myriad of discussions surrounding privacy, compliance, technical measures and regulations, the customer (user, individual) is the ultimate stakeholder, whose data comprises the crux of the issue.

\vspace{5pt}
\fbox{
\begin{minipage}{24em}
RQ2: The implementation of technical measures for data privacy compliance involves many roles and responsibilities, most notably those of the so-called \say{Go-Betweens}.
\end{minipage}}

\subsection{Interactions (RQ 3)}
\label{interactions}
There exists a series of crucial interactions that must take place for technical measures to be realized. Proposing that structuring these is a key first step to understanding the technical measures themselves, we distinguish three particular interaction types. As such, our main focus lies in the interactions where technical forces are at play, i.e. where the technical measures are ultimately implemented.

\subsubsection{Technical-Technical} 
Before the influence of legal support even comes into play, much
of the interaction regarding the technical measures for privacy compliance occurs within the technical sphere. It is in these technical-technical interactions that many of the technical roles described above operate in their daily capacities.

The first set of such interactions occur \say{vertically} within the engineering side of compliance. A privacy engineer may not be \say{working directly with legal to do things, [but] still taking direction from [an] engineering manager} (I1-T). This is validated from the statement of another privacy engineer: \say{Sometimes I'll interact with a program manager or a technical program manager, if there is a need, like maybe they have broader insight into a project that a given engineer doesn't.} (I4-T). Privacy architects involved also have a \say{factual reporting line to the product management} (I2-T). In this way, one begins to see that much of the guidance and direction for technical decisions regarding privacy is handed down from direct managers within an engineering vertical. Specifically, \say{the head of the tech vertical will always get down to the cut and dry for us, which is directly asking Legal questions.} (I1-T).

Of more of the \say{horizontal} nature come interactions between privacy engineers and other parties. One privacy engineer is \say{in almost constant communication with either subject matter experts, software engineers, or both, because sometimes they overlap} (I4-T). The dialogue with these subject matter experts is important for privacy engineers to stay up-to-date on relevant topics, both technical and legal. Interactions with software engineers can be viewed as more \say{diagonally down}, as privacy engineers provide the necessary guidance and policy for privacy-protecting systems. 

A final important set of interactions within the technical sphere is interactions with peers within the same group or team. An illustrative example of these came from the interview with I4-T, who described a team of many privacy engineers, all of whom were specialized in something slightly different. Because of this, each member's strengths could be drawn upon for the mutual benefit of all privacy engineers. These horizontal relationships, therefore, play a vital role as well. 

\subsubsection{Technical-Legal} As introduced in previous sections, the need to implement technical measures to comply to privacy regulations has brought together two inherently different fields, and more abstractly, two quite different ways of thinking. 

Starting from the non-management level of the engineering vertical, the main technical-legal interactions occur with the appointed DPO. In the words of an architect, \say{Sometimes we interact with the data protection officer, when it's about the specific interpretation of legal requirements or validating a certain technology fits to the requirement or not.} (I2-T). This description is very insightful in the way that it binds the interpretation of legal requirements to the validation of a technology. Precisely this illuminates the go-between role of the DPO. Another source of legal support for engineers, going back to the horizontal interactions described above, was described by I4-T: \say{Who do I go to with a legal question? Normally, I'll ask senior experts, people who are privacy engineers, who are former lawyers, it's probably my first pitstop.}

Viewing this category from more of the legal-technical direction, the interaction between legal support (besides the DPO) and non-management technical parties seems to be of the rarest occurrences. I8-L mentions sometimes liaising with an organization's structured IT or development teams, claiming \say{I can't imagine how I would be able to actually do my work without kind of a real input from the IT team.} (I8-L). I16-L also mentions such interactions.

A set of interactions that takes place undoubtedly more often is the dialogue between legal teams and the technical leadership of organizations. One illustration of the technical-legal relationship from the legal perspective is as such: \say{We have a designated Information Security Officer, who works with these issues from the IT side. And usually I share my findings with him, and he shares his findings with me. So we're kind of communicating really, I would say, really well.} (I8-L). A similar level of communication is conveyed from an external DPO, who initially sets up an \say{at least two hour interview with the IT leader of the companies} (I12-L).

The final type of technical-legal interactions comes in the form of \textit{cross-teams}, or \textit{cross-functional teams}, a topic which came up in some interviews. These teams typically consist of members from different departments, including those that are more technically or legally oriented. While such a concept does not always exist, it describes an interesting avenue for cross-disciplinary exchange.

\subsubsection{Legal-Legal} 
A final category of interactions described in some interviews were those of a purely legal nature. Since these are out of scope of this paper, they are only mentioned briefly.

Outside counsel or even an external DPO may often work with the legal compliance team within an organization, serving a more advisory or guiding role. Similarly, an internal DPO may seek consultation from an external consultant.

An interesting example comes out of the interview with I14-L, with the idea of a \say{Data Governance Council}. While this is not a purely legal team, it is placed under the legal-legal category. Specifically, this council \say{is more of an independent board of senior leadership... if anything ever needs to be flagged from both business and clients [on one side] and legal [on the other], it gets escalated to them, and they can make more of the business related approaches to risk} (I14-L). This is certainly an interesting sub-structure, but something that is not widely adopted, as far as can be determined from the interview findings.

\subsubsection{On \textit{Better and/or More Interaction}}
Regarding the question of whether better interaction between technical and legal roles is needed, the responses represent a particularly crucial point of analysis, especially in the scope of RQ3. Many interviewees agree that improved interaction (in frequency or quality) between the technical and legal sides of data privacy compliance would indeed be beneficial. Some interviewees, such as I8-L, note that day-to-day communication may not be necessary, but \say{it does require processes to have been built out}, once again supporting the importance of the process behind technical measures. One of the most concrete responses to the question originates from I14-L, who illustrates the following dynamic:
\begin{quote}
    \say{Yes, definitely. The one way I see it is tech is driving a lot of the new changes in privacy. And it's not enough for it to just be a legal focus team without consulting with tech, I think it needs to be a mix of the two, because the two kind of balance each other out, because that's what privacy is becoming.}
\end{quote}
In this way, the interdisciplinary nature of technical measures, and the process of their implementation, is highlighted as truly interdisciplinary. A dissenting opinion, though, is expressed by I4-T, who adamantly responds: \say{No, they're not going to know how to do what I do.} To a lesser degree, I5-LT also agrees with this sentiment, but for the reason that positions such as the DPO exist precisely to bridge the technical-legal gap. As such, it is incredibly interesting that these two opposing views come from the Go-Between roles.

\vspace{5pt}
\fbox{
\begin{minipage}{24em}
RQ3: For technical measures to be implemented, an intricate web of interactions must take place, especially those of an interdisciplinary nature.
\end{minipage}}

\section{Discussion}
\label{discuss}

\subsection{Breaking Down Technical Measures}
The findings presented in Section \ref{rq1} clearly illustrate the multi-faceted nature of the implementation of technical measures, particularly in its perception by different roles involved in the privacy compliance process. Breaking down these responses illuminates some interesting insights, which ultimately points to a divergence in technical and legal perspectives on the matter, as well as the inherent complexity of the notion of technical measures.

Analyzing the discussion of technical measures in the interviews, one can compartmentalize the practical understanding thereof in a somewhat clear way. Firstly, technical measures thought from the purely technical point of view cannot be ignored. In a perhaps surprising way, this straightforward approach was preferred by legal interviewees, pointing to specific technologies that are referenced in GDPR, for example. Another predominant viewpoint comes with the understanding of technical measures as a risk assessment. In other interviews, the idea of data flow was seen as paramount to understanding technical measures. In these ways, the notion of technical measures is made concrete, at least in the eyes of the privacy professional being interviewed.

A matter that is left less concrete is the role of Privacy-Enhancing Technologies (PETs), something with which technical interviewees provided valuable insight. Although one would be hard pressed to refute the privacy-preserving capabilities of such technologies, their practical prevalence as technical measures remains in question due to factors such as complexity, education, and legal interpretability.

The flip side must also be analyzed. In an equally surprising manner, technical interviewees tended not to focus on specific technologies that may suffice as technical measures. Rather, they were more inclined to discuss broader issues surrounding technical measures, the understanding thereof, and the overall compliance process. The vagueness of technical measures and the interpretation required because of it was an important talking point, one that was also promoted by legal voices. The discrepancy between legal mandate and technical implementation was likewise emphasized.

Unexpectedly, interviewees did not generally show awareness about tools supporting the implementation of technical measures for data privacy compliance. 
One of the reasons for low awareness and lacking adoption of such tools in practice can be attributed to insufficient attention to both technical and legal perspectives.

Uniting the two camps is the general characteristic shared by most responses that the process of implementing technical measures is not a codified science. There is much room for interpretation, and concurrently, there exist many factors beyond the technology itself. Ultimately, one can hypothesize that such factors lead precisely to the differences in viewpoints over technical measures reflected in the interviews. Such a finding confirms the unique nature of technical measures, as well as demands deeper inquiry into its understanding.

\subsection{Lack of Structural Uniformity}
A cursory analysis of the findings presented in \ref{rq2} reveals that there is a multitude of roles involved in the privacy compliance process. By studying such roles (and their responsibilities), one can begin to understand the technical measures in question beyond the measures themselves, by viewing how the ecosystem behind them exists and functions.

The role of legal can be characterized by its advisory and supporting nature. This can be extracted from the diction of legal interviews with frequent language revolving around \textit{supporting}, \textit{ensuring}, and \textit{interpreting}. An interesting junction in the legal side of privacy compliance comes with the overlapping roles of in-house and external counsel, legal consultants, and compliance teams. While the utilization of such roles vary by organization size and data processing activity, it is seemingly the case that the responsibilities amongst these roles can be easily conflated. This may call for a better understanding of and distinction between the legal roles involved in the implementation of technical measures.

The technical influence in the interpretation and implementation of technical measures must also be stressed. Indeed, one can observe that much of the work surrounding technical measures, including the interpretation of regulation, may exist solely in the engineering vertical. This emphasizes foremost the crucial part played by management roles within the engineering vertical, who ultimately are responsible for overseeing the implementation of technical measures.

This still does not fully exclude a necessity for the involvement of legal experts. In some organizations, engineers themselves have a direct or indirect line (e.g. through management) of communication to escalate issues or questions directly to legal teams.

Serving as the necessary bridge between the technical and legal sphere of privacy compliance, the term of \textit{"Go-Between"} has been coined. These roles are incredibly interesting to analyze in the way that inherent to the job is the connection of two distinctly different fields of thought. In this way, two important connectors have surfaced, one that clearly \say{leans} more technically, and one more legally. The role of Privacy Engineering as the driving factor for an organization's policy on privacy regarding technical measures becomes very important. Analogously, the Data Protection Officer must balance legal know-how with a cursory understanding of the systems in play. Above all, the interaction of these Go-Betweens, both among themselves and with their counterparts, is presumably of utmost importance, yet these interactions remain not well studied. This becomes the building block for the ensuing discussion. 

As such, the number of roles that in some way tangent upon the implementation of technical measures is quite sizeable. One possible analysis of this fact may realize the opportunity for great collaboration amongst this variety of roles. A more skeptical observer may posit that such a disparity in the roles involved may reflect that there exists no structured and widely adopted model from an organizational perspective for approaching privacy compliance. In fact, we can assume that while both technical and legal perspectives have their own approaches to the implementation of technical measures, there is a lack of exchange and integration between them. We also see that the introduction of "Go-Betweens" is an effort to compensate such disparity through the introduction of a role that is interdisciplinary by nature. 

\subsection{Interdisciplinarity}
The findings of RQ3 highlight the prevalence of interactions between roles of differing expertise. This is made clear in the technical-legal category, but is also true for the interactions occurring \textit{within} the technical vertical. In this way, the interdisciplinarity of privacy compliance, centered around technical measures, becomes evident.

It is for this reason that the interactions surrounding the implementation of technical measures become interesting points of investigation. With such dynamism and interconnection, there is bound to be fruitful collaboration, but also undoubtedly inefficiencies and challenges. Conducting a deeper dive into these intersections would merit worthwhile future work, in order to understand better the complexities of such dynamic interaction.

Our findings indicate a clear agreement for better and/or more interaction between the various roles introduced here, particularly between the technical and legal ones. It is in these interdisciplinary relationships that a clearer, more unified understanding of technical measures can be achieved, so as to harmonize the inherently different perspectives of the two. While the manifestation of such interdisciplinarity may not be so straightforward in practice, it becomes an excellent standard for which to strive.

\section{Threats to validity}
\label{sec:threats}
\textbf{External validity}: Since this is exploratory research investigating practitioners, we placed a high emphasis on external validity according to Cook and Campbell~\cite{Campbell2015,Cook1979}. To achieve this, we followed two strategies: First, we targeted expert participants with significant experience as privacy professionals in the industrial context. This ensured that reported issues have practical relevance and were not due to the lack of experience. Second, we ensured heterogeneity among participants by recruiting them from different industries and domains. As for the \textit{transferability} of the observations, we acknowledge the size of our sample remains small, and we cannot generalize the findings to all the organizations. Nevertheless, we reduced this threat by interviewing practitioners in different roles from different domains and differently sized companies. Moreover, we observed a convergence of the findings during the interviews and analysis. We assume in another similar context, the observations will be similar. Nonetheless, our study results require further confirmation by subsequent investigations.

\textbf{Construct Validity}: Since our study is exploratory, construct validity was not the primary concern. Nevertheless, we minimized threats to it by (i) piloting and reviewing the interview guideline among researchers, (ii) performing each interview via video call, allowing for immediate clarification of questions, and (iii) conducting some interviews with at least two researchers, one guiding the conversation, one taking notes and asking for clarification.

\textbf{Reliability}: Reliability of outcomes was important for our investigation. We achieved it by adhering to the GT (Grounded Theory) approach. We rigorously protocolled each step and followed structured coding procedures to develop a theory based on the raw data obtained in the interviews.


\section{Conclusion}
\label{sec:conclusion}
In this work, we investigated the current understanding of the term "technical measures" among privacy professionals and the processes that take place to implement these measures in practice, focusing on the roles and interactions of the involved parties in both technical and legal sectors. Our results demonstrate that there is no clear mutual understanding and commonly accepted approach to handling the process of implementation of technical measures.

From a practical perspective, our results demonstrate the importance of the interaction between legal and engineering functions for the implementation of technical measures. Still, such interactions remain rare and transpire in various forms. Hence, further enhancement of such interactions is a clear practical step towards improving the implementation of technical measures for compliance.
From a scientific perspective, our results support previous suggestions about the need for closer interactions between software engineering and legal roles in regulatory compliance. Our results suggest that it is important to consider data privacy compliance practices in the industry to assure the practical relevance of research results.

In our opinion, future work in this direction should be focused on a closer empirical research of technical-legal interaction required for the implementation of technical measures, investigation of barriers to the development and practical adoption of tools and technologies that support the implementation of technical measures, as well as on the education and awareness of practitioners about the existing tools and technologies such as PETs.
As mentioned in Section \ref{discuss}, the understanding of PETs as technical measures and how they can be rectified with regulatory requirements must be further investigated. Moreover, although the study subjects were senior practitioners, the sample size was relatively small and thus, generalizations can be difficult. Thus, future work should aim to collect further evidence to refine and support the observations described in this paper.

Another pertinent avenue for future work should focus on the people, roles, and responsibilities involved in the process of privacy compliance. Specifically, how issues such as privacy education and awareness can affect the implementation of technical measures within organizations may prove to be a highly interesting point of investigation. A pressing question then becomes: how can technical measures be understood comprehensively when the structures behind them are so disparate? This forms the basis of future inquiry.

The common ground for these future avenues is founded in the interdisciplinary relationships that became a predominant theme in the findings, and subsequently, this paper. It is these interactions, we argue, where the process of implementation of technical measures (and the understanding thereof) comes to a crucial junction point.

\newpage

\begin{acks}
This work has been supported by the German Federal Ministry of Education and Research (BMBF) Software Campus grant LACE 01IS17049 and the Bavarian Research Institute for Digital Transformation (bidt) funding for projects "Differential Privacy: A New Approach to Handling Social Big Data" and "Coding Public Value".
\end{acks}

\bibliographystyle{ACM-Reference-Format}
\bibliography{main}

\end{document}